\documentstyle[12pt]{article}
\input epsf.tex

\newcommand{\nc}{\newcommand}
\nc{\beq}{\begin{equation}}
\nc{\eeq}{\end{equation}}
\nc{\beqa}{\begin{eqnarray}}
\nc{\eeqa}{\end{eqnarray}}
\nc{\lra}{\leftrightarrow}
\def\sfrac#1#2{{\textstyle{#1\over #2}}}
\nc{\sss}{\scriptscriptstyle}
{\nc{\lsim}{\mbox{\raisebox{-.6ex}{~$\stackrel{<}{\sim}$~}}}
{\nc{\gsim}{\mbox{\raisebox{-.6ex}{~$\stackrel{>}{\sim}$~}}}

\def\eps{\epsilon}
\def\sL{{\!\sss\Lambda}}
\def\bL{\bar\Lambda}

\begin{document}

\begin{titlepage}
%%%%%%%%%%%%%%%%%%%%%%%%%%%%%%%%%%%%%%%%%%%%%%%%%%%%%%%%%%%%%%%%%%%%%%%%%%%%%%%
\begin{flushright}
McGill 00-35\\
hep-ph/0012090\\
\end{flushright}

\vskip.5cm
\begin{center}
{\Large{\bf A Small Cosmological Constant \\
 \vskip0.4cm
from Warped Compactification with Branes}}
\end{center}
\vskip1.5cm

\centerline{ James M.\ Cline and Hassan Firouzjahi}

\centerline{Physics Department, McGill University,
Montr\'eal, Qu\'ebec, Canada H3A 2T8}

\begin{abstract}
\vskip 3pt

We present a possible explanation for the smallness of the observed
cosmological constant using a variant of the
Randall-Sundrum(RS)-Goldberger-Wise paradigm for a warped extra dimension.  
In contrast to RS, we imagine that we are living on the positive tension
Planck brane, or on a zero-tension TeV brane.  In our model there are two
solutions for the scalar field in the bulk and the corresponding brane
separations, one of which is tuned to have zero cosmological constant.  
We show that in the other solution, which is a false vacuum state, the 4-D
cosmological constant can be naturally small, due to exponential
suppression by the warp factor.  The radion is in the milli-eV mass range,
and if we live on a TeV brane its couplings are large enough that it can
measurably alter the gravitational force at submillimeter distances.  In
this case the Kaluza-Klein excitations of the graviton can also contribute
to submillimeter deviations from Newtonian gravity, and we have in
addition the collider phenomenology of the usual TeV-scale radion.

\end{abstract}

\vfill
\end{titlepage}

{\bf 1. Introduction.} There is mounting evidence that we live in a
universe with a vacuum energy density $\Lambda$ which is about 0.7 of the
critical density \cite{SN}.  This value is close to 120 orders of
magnitude less than the Planck density, $M_p^4$, which would be the
natural expectation for the size of $\Lambda$ from quantum field theory.  
If there exists some mechanism to make $\Lambda$ small, it would seem much
simpler if its value was zero than $10^{-120} M_p^4$.  Our motivation in
this letter is to present a possible explanation for this enormous
hierarchy of scales, taking advantage of recent ideas involving 3-branes
embedded in an extra dimension.

A similar hierarchy problem, though much less severe, is that of the weak
scale (100 GeV) versus the (reduced) Planck scale, $M_p = (8\pi
G_N)^{-1/2} = 2.43\times 10^{18}$ GeV.  Randall and Sundrum (RS) presented
an interesting way of explaining this small ratio using two 3-branes in a
5-D anti-deSitter space with a compact extra dimension
\cite{RS}-\cite{RS2}.  It is tempting to look for a similar application of
the RS idea for the cosmological constant problem; several attempts to use
one or more extra dimensions to get a small $\Lambda$ have recently been
made \cite{self-tuning}-\cite{others}.  In this paper we will suggest that
the RS idea can, with only minor changes, account for the smallness of the
observed $\Lambda$.  Our model relies heavily on stabilization of the
distance between two branes using a bulk scalar field, as was suggested by
Goldberger and Wise \cite{GW2}.

Our explanation only partially addresses the cosmological constant
problem, in that we assume there is some unknown mechanism which forces
the ground state of the universe to have vanishing 4-D vacuum energy.
We show that there can be a false vacuum state with $\Lambda$ nonzero and
exponentially small, as illustrated in figure 1.  Its size is determined
by the separation between the two branes, which in turn depends on the
potential of the bulk scalar field $\phi$.  The degree of freedom which is
varying between the true and false vacua is the {\it radion}, the
dynamical field associated with fluctuations in the size of the extra
dimension.

To set the stage, we will be working in a 5-D theory with the
extra-dimensional coordinate $r$, whose metric is parametrized by 
\beq
   ds^2 = e^{2A(r)}\left(dt^2 - e^{2\sqrt{\bL}t}\sum_i dx_i^2\right) -
dr^2.
\eeq
We note that slices of constant $r$ represent 4-D deSitter spaces with
vacuum energy $\Lambda = 3\bL/(8\pi G)$, where $G$ is the 4-D Newton
constant.
There are two 3-branes, located $r=0$ and $r=r_1$ respectively.
There is also a scalar field $\phi$ with potential $V(\phi)$ in the bulk,
and separate potentials $\lambda_0(\phi)$ and $\lambda_1(\phi)$ on the
two branes.  $V(\phi)$ contains a negative bulk cosmological constant
which induces the behavior $A(r)\propto -r$, so that $e^{2A(r)}$ is
exponentially decreasing.  Using notation similar to that of reference
\cite{DeWolfe}, the action is 
\beq
\label{action}
	S = \int d^{\,5}x \sqrt{g}\left(-\frac1{4\kappa^2} R + \frac12
(\nabla\phi)^2
	-V(\phi)- \lambda_0(\phi) \delta(r) -
	\lambda_1(\phi)\delta(r-r_1)\right)
\eeq
Here $\kappa^2 =\frac{1}{2} M^{-3}$ defines the 5-D gravity scale
$M$, which is assumed to be of order $M_p$.
To simplify the analysis, we will take the brane
potentials to have the form
\beq
\label{bps}
	\lambda_i(\phi) = T_i + \gamma_i(\phi^2 - \phi_i^2)^2
\eeq
with $\gamma_i\to\infty$, {\it i.e.,} the brane potentials are
stiff.  This will make the boundary conditions for the scalar field
simply
\beq
	\phi(0) = \phi_0,\qquad \phi(r_1) = \phi_1,
\eeq
and the potentials will have the values $\lambda_i(\phi_i) = T_i$, where
$T_i$ denotes the tension of the respective brane.  

In the remainder we will show that for the same simple choice of bulk
potential used in the exact solution of ref.\ \cite{DeWolfe},
it is possible to find a metastable solution to the equations of motion
whose false vacuum energy (in the 4-D effective theory) is exponentially
close to that of the true vacuum.  By assuming the latter has vanishing
cosmological constant, we can thus explain the small observed value of
$\Lambda$ in our universe by imagining that we are stuck in the false
vacuum.
\bigskip

{\bf 2. Superpotential method.}  To construct solutions to the
coupled equations for 5-D gravity and the scalar field, we will take
advantage of the superpotential method discussed in ref.\ \cite{DeWolfe}.
One introduces a superpotential $W(r)$ and a function 
\beq
\gamma(r) = \sqrt{1 + {9\bL\over \kappa^4W(r)^2}\, e^{-2A(r)}}
\eeq
such that the desired potential $V(\phi)$ is given by
\beq
\label{Veq}
	V(\phi(r)) = {1\over 8\gamma^2} \left({\partial W\over \partial
		\phi}
	\right)^2 - \frac{\kappa^2}{3} W^2.
\eeq
Then solutions to the coupled equations for $A$ and $\phi$ can be generated
from the first order equations
\beq
\label{eom}
	A' = -\frac {\kappa^2}{3} W \gamma;\qquad \phi' = {1\over 2\gamma}
	{\partial W\over \partial \phi} \,.
\eeq
For flat branes (when $\Lambda=0$), $W$ can be regarded as a function of
$\phi$ alone, but for bent branes ($\Lambda\neq 0$), it must be considered
as a function of $r$, and in that case one interprets ${\partial W\over
\partial\phi} = {1\over\phi'}{dW\over dr}$.   The boundary conditions 
at the brane positions $r=0$ and $r=r_1$ are 
\beqa
\label{bc0}
	\lambda_0(\phi_0) &=& W\gamma|_{r=0\phantom{_1}};\qquad
	\phi(0) = \pm\phi_0
	\\
\label{bc1}
	\lambda_1(\phi_1) &=& -W\gamma|_{r=r_1};\qquad \phi(r_1) = 
\pm\phi_1
\eeqa

{\bf 3. Solutions with vanishing $\mathbf\Lambda$.}  
Let us first present a ground state solution in which the branes are flat
and the 4-D cosmological constant $\Lambda$ is tuned to be zero, hence
$\gamma(r) = 1$.  We will
later perturb this solution with a small positive value of $\Lambda$.
The unperturbed functional form for the superpotential is taken to be
\beq
\label{Weq}
	W_0(\phi) = {3\over \kappa^2 L}  -  b\phi^2,
\eeq
where $L$ is a length scale that turns out to be the curvature radius of
the 5-D
anti-deSitter space in the limit where the back reaction of the scalar
on the geometry is small.  The resulting scalar potential is simply a
polynomial in $\phi$ of degree 4.  It is easy
to integrate the equations of motion (\ref{eom}) to obtain the solutions
\cite{DeWolfe}
\beqa
	\phi &=& \phi_0\, e^{-br}\nonumber\\
	A &=& \hat{A}-{r\over L}
	-\frac{\kappa^2}6 \phi_0^2\, e^{-2br},
\eeqa
where $\hat{A}$ is a constant of integration (we have already chosen 
$\phi_0$ so as to satisfy the boundary condition on $\phi$ at $r=0$.)
The value of $\hat{A}$
is physically irrelevant, so we can for convenience choose it such that
$A(0) = 0$.  Applying the boundary condition on $\phi$ at the second
brane determines $r_1$, the size of the extra dimension:
\beq
	e^{-br_1} = {\phi_1 \over \phi_0}
\eeq
This solution is essentially the same as the ones found in references
\cite{GW2} and in particular \cite{DeWolfe}, differing from the latter
only 
by our choice of stiff brane potentials.  The stability of this solution
against small perturbations has been demonstrated in \cite{CGK}.
The large hierarchy of energy scales between the two branes is generated
by assuming that $b$ is parametrically small compared to $1/L$.  In this
case, although $e^{-br_1} \lsim 1$, the warp factor will be
exponentially small, $e^{-r_1/L}\ll 1$.

The remaining boundary conditions are satisfied by the constraints
\beqa
\label{Teq0}
	T_0 &=& {3\over \kappa^2L}  - b\phi_0^2 \\
\label{Teq1}
	T_1 &=& -{3\over \kappa^2L} +  b\phi_1^2
\eeqa
on the brane tensions.  The fact that we have two fine-tunings instead of
the one which is expected on the basis of counting parameters
\cite{DeWolfe} is due to our choice of superpotential $W_0$.  There is
actually a family of superpotentials giving rise to the same physical
potential, and these can be found by integrating the differential equation
(\ref{Veq}).  A new constant of integration would then arise, which we
have already fixed by our choice of $W_0$.  If the brane tensions were not
related to each other as prescribed by (\ref{Teq0}-\ref{Teq1}) we would be
forced to find the appropriate form of $W_0$.  Thus there is only one
essential fine-tuning, and this is the one that corresponds to setting the
4-D cosmological constant $\Lambda$ to zero.

It is important to notice that, in addition to the solution at finite
brane separation, there is another one with the second brane at
$r=\infty$.  In other words, the second brane is simply removed from the
picture.  The extra boundary conditions associated with the second brane
are no longer relevant when it is at infinity since (due to the vanishing
warp factor) it makes no contribution to the action.
\bigskip

{\bf 4. Nearby solutions with nonzero $\mathbf\Lambda$.}  Let us suppose
that the tuning of brane tension $T_0$, eq.\ (\ref{Teq0}), is enforced, so
that the solution where the second brane removed still exists, with
vanishing 4-D cosmological constant.  Consider what happens to the other
solution if the tension $T_1$ is no longer tuned according to
condition (\ref{Teq1}).  We expect that $\Lambda$ is no longer zero in
this case.  The equations of motion with nonzero $\Lambda$ are difficult
to solve exactly, so we will instead solve them perturbatively, to first
order in $\Lambda$.  

The first step in this procedure is to realize that, although $V(\phi)$
should still have the same functional form as in the $\Lambda=0$
solutions, the superpotential need not be the same.  In general, it must 
be corrected in such a way that $V(\phi)$ is unchanged.  Let us denote
the new solutions and superpotential by
\beqa
	\phi_\sL &\cong& \phi + \delta\phi;\qquad A_\sL \cong A +\delta A
	\nonumber \\
	W_\sL(r) &\cong& W(\phi_\sL(r)) + \delta W(r)
\eeqa
where the quantities $\phi(r)$ and $A(r)$ refer to the bulk solutions with
$\Lambda=0$ found in the previous section, and $W(\phi)$ is the
corresponding superpotential (\ref{Weq}).  By taking the variation of
eq.\ (\ref{Veq}) and keeping terms which are first order in $\Lambda$ we
get a differential equation for $\delta W$,
\beq
	\delta W' - \frac{4}{3}\kappa^2 W\delta W = \frac{9}{4\kappa^4} \bL
        e^{-2A}	\left({1\over W}\, {\partial W\over \partial \phi}\right)^2
\eeq
which by using the equations of motions (\ref{eom}) can be written as
\beq
	(e^{4A} \delta W)' = \bL e^{2A}\left(\phi'\over
A'\right)^2.
\eeq
This can be formally integrated to solve for $\delta W(r)$,
\beq
\label{dWeq}
	\delta W = \bL e^{-4A}\int_0^r e^{2A}\left(\phi'\over
	A'\right)^2 dr + C e^{-4A}
\eeq
To determine the constant of integration $C$, we should impose the
boundary condition (\ref{bc0}) on the tension of the first brane.  
Recall that the value $T_0$ was already fixed by demanding the existence
of the solution with only one brane and $\Lambda=0$.  Since the brane 
potentials are assumed to be stiff, $\lambda_0(\phi_0) = T_0$ regardless
of whether $\Lambda=0$ or not.  This leads to the requirement that the
variation of eq.\ (\ref{bc0}) vanishes:
\beq
	\delta W(0) + {9\bL e^{-2A(0)}\over 2\kappa^4
W(\phi_0)} = 0.
\eeq
Recalling the definition $A(0) = 0$, this fixes $C$ to be
\beq
\label{Ceq}
	C = -{9\bL\over 2\kappa^4 W(\phi_0)}	
\eeq
In fact, because of our assumption that $b \ll 1/L$, needed to get
a large hierarchy of scales, the $C$ term is the dominant one in $\delta W$.
The first term in (\ref{dWeq}) is of order $\phi'^2 \sim b^2$ and can 
therefore be neglected.

Once $\delta W$ is known, it is straightforward to find the perturbations
$\delta\phi$ and $\delta A$.  In particular, the equation for $\delta\phi$
is
\beq
	\delta\phi' = {\delta W'\over 2\phi'} - {9\bL e^{-2A}\over 2\kappa^4
	W(\phi)^2} \phi' + {1\over 2} {\partial^2 W\over \partial
	\phi^2} \delta\phi
\eeq
Again, since $b\ll 1/L$, the first term dominates.  We thus find the 
approximate solution
\beq
\label{dphi}
	\delta\phi \cong  {3\bL L\over 4\kappa^2 b\phi_0}\left(
	e^{4r/L}-e^{-br}\right).
\eeq
To obtain this result, we have assumed that $1/L \gg b, 
b\phi^2$.  (The fact that $b\phi_0$ appears in the denominator of
this result might seem to upset our perturbation expansion; we will
address this potential problem presently.)  
The constant of integration was chosen so that
the boundary condition $\phi(0) = \phi_0$ is respected by the perturbed
solution.  

All that remains is for us to impose the two boundary conditions at
the second brane.  These will determine the new position of this brane,
$r_{\sL}$, and the 4-D cosmological constant $\Lambda$.  The b.c.\ on
$\phi$ is 
\beq
\label{phibc}
	\phi_1 =  \phi_0\, e^{-b r_{\sL}} + 
	{3\bL L\over 4\kappa^2 b\phi_0}\left(
        e^{4r_{\sL}/L}-1\right)
\eeq
and the condition for the brane tension is
\beq
	T_1 = -W(\phi_1)\left(1 + {9\bL e^{-2A(r_{\sL})}\over 2
	\kappa^4W(\phi_1)^2}\right) - \delta W(r_{\sL})
\eeq
Using the hierarchy $e^{-4A(r_\sL)} \gg e^{-2A(r_\sL)} \gg 1$,
this can be approximately
solved for $\bL$ to give
\beq
\label{mainresult}
	 \bL \equiv {8\pi G\over 3}\Lambda \cong \frac{2\kappa^2}{3L}\, e^{-4r_         \sL/L} (T_1 + W(\phi_1) ) 
\eeq
This is our main result: if the interbrane distance is around  $r_\sL\sim
\ln(10^{30})L \sim 70L$, then the physical 4-D cosmological constant can
be the observed value even if $(T_1 + W(\phi_1)$ is of order the Planck
energy density.  (Recall that $T_1 = -W(\phi_1)$ is the fine-tuned value
that gave us $\Lambda=0$ in the previous solutions.) To see if such a value
of $r_\sL$ is natural, we must solve the $\phi$ boundary condition
(\ref{phibc}):
\beq
\label{rsLeq}
	e^{-b r_{\sL}} = {\phi_1 \over \phi_0}
	- {\left(T_1 + W(\phi_1)\right)\over  2 b\phi_0^2}
\eeq
However we must consider the fact that the second
term on the right hand side of this equation is of order $1/b$.  
A small amount of tuning is required so that 
$(T_1 + W(\phi_1)$ is of order $b^2$, so that the new term can be 
treated perturbatively.    Once this is done however, there
is no great difficulty in achieving the enormous hierarchy of 120 orders of
magnitude for $\Lambda$.  For example if $bL = 0.01$, the r.h.s.\ of
(\ref{rsLeq}) need only be as small as 0.5 to achieve the required
separation of $r_\sL = 70L$.  We have therefore succeeded ameliorating
the cosmological constant hierarchy problem in the same way as the 
original Randall-Sundrum model addressed that of the weak scale.
The mild fine-tuning we demanded for the second brane tension
might be merely a technical requirement for the convenience of
analytically demonstrating the mechanism.  Possibly it still works
even without this small amount of tuning, though the solutions are harder
to find in that case.

Recently ref.\ \cite{GKL} pointed out that any warped compactification
solution involving a scalar field must satisfy certain consistency
conditions, notably that $\sum_i T_i + \int \phi'^2 dr \cong 0$ in the
case where $\Lambda \ll M_p^4$.  The present solution has already been
shown in \cite{GKL} to satisfy this relation when $\Lambda =0$, so we
should verify that our perturbed solution also satisfies it.  To first
order in the perturbation, we should find that $\delta T_1 + 2\int \phi'
\delta\phi' dr = 0$, where $\delta T_1$ is the mistuning of $T_1$ away
from its $\Lambda=0$ value, $\delta T_1 = T_1 + W(\phi_1)$.  It is
straightforward to show that in the regime where $b\ll 1/L$,
$2\int_0^{r_1} \phi' \delta\phi' dr = \delta W|^{r_1}_0 \cong C
e^{-4A{r_1}}$, and using (\ref{Ceq}) and (\ref{mainresult}) that the
consistency condition is satisfied.

To complete our argument we should demonstrate
that the 4-D Newton constant $G$ can take its observed value without
requiring
any fine tuning.  By integrating over the extra dimension in the 5-D
action (\ref{action}) to obtain the 4-D effective action, one finds the
relation
\beq
	{1\over 16\pi G} = {1\over 4\kappa^2}\int_{-r_\sL}^{r_\sL} dr\, e^{2A}
	\cong {L\over 4\kappa^2}
\eeq
Thus it is natural to assume that $G$, $\kappa^2$ and $L$ are all of
the order $M_p$ to the appropriate power, and no additional tuning 
is needed to localize gravity.

Although we have succeeded in constructing the approximate solution
corresponding to the local minimum of the radion potential, one might
wonder why we are not able to find the unstable solution at the local
maximum.  Evidently this is nonperturbative in $\Lambda$.  Recall that we
needed to do some mild tuning, $T_1+W(\phi_1)= O(b^2)$ to keep our
perturbation expansion under control.  Yet the unstable solution exists
even when $T_1+W(\phi_1)$ is exactly zero.  It thus seems plausible that
the barrier height is large, although we expect in order of magnitude that
its size is governed by the same warp factor $e^{-r_\sL/L}$ which
suppresses the perturbative solution value.

\centerline{\epsfxsize=3.5in\epsfbox{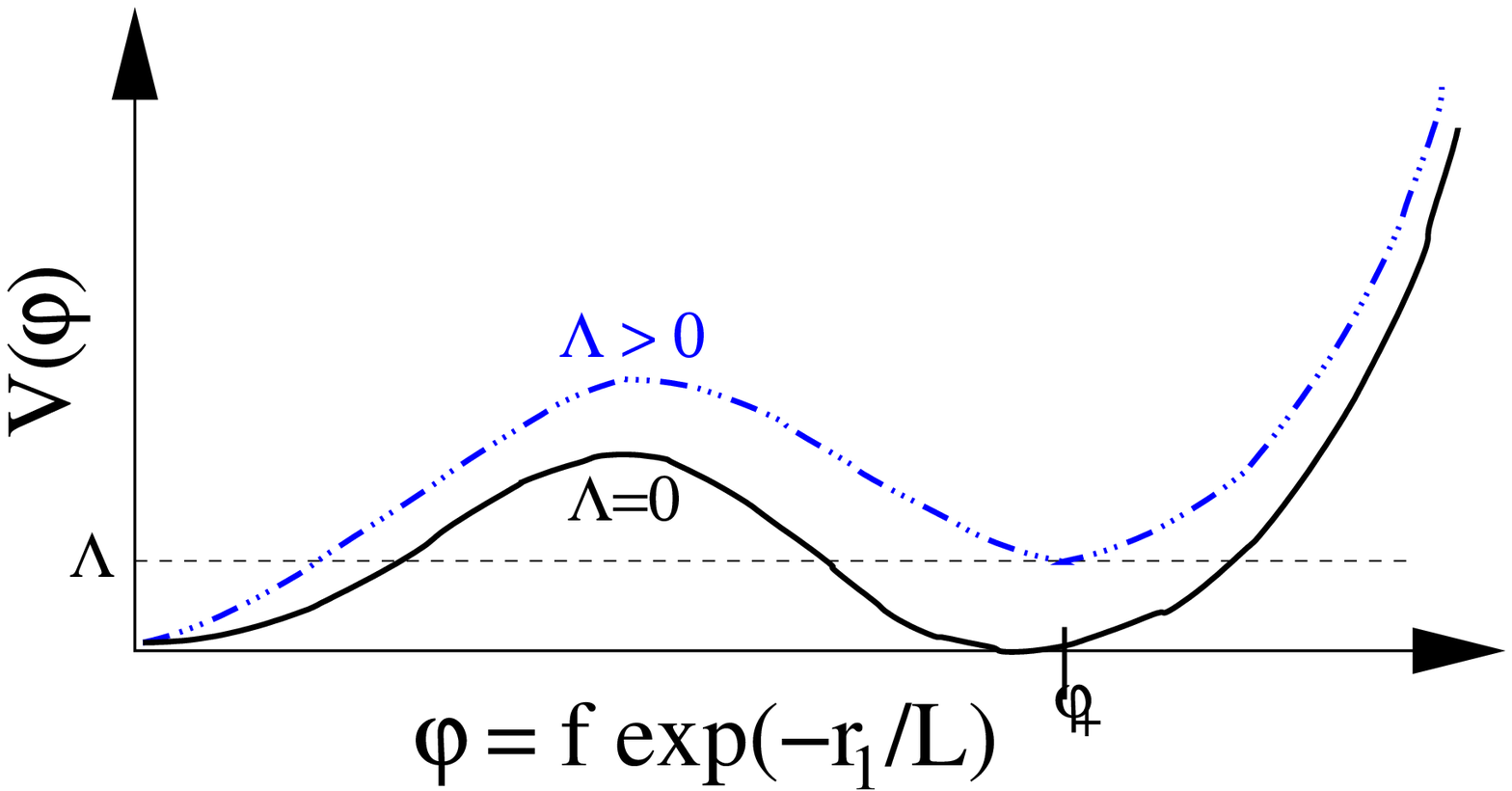}}
\vspace{0.in}
\noindent {\small
Figure 1: Qualitative form of the radion potential,
for the two cases $\Lambda>0$ and $\Lambda=0$.  The
false vacuum state at $\varphi=\varphi_+$ 
naturally has $\Lambda\sim 10^{-120} M_p^4$.}
\bigskip

{\bf 5.\ Lifetime of the false vacuum.} Our proposal is only viable if the
false vacuum has a sufficiently large lifetime that we could still be in
it at the present time.  To study this we should construct the bounce
solution of the Euclideanized radion action and check that the lifetime
$\tau \sim \Lambda^{-1/4} e^{S_b}$ exceeds the age of the universe, where
$S_b$ is the bounce action and $\Lambda^{-1/4}\sim 10^{-4}$ eV is the
typical energy scale that will appear in the prefactor in the saddle point
approximation for the path integral for the rate of false vacuum decay.
We will take the following ansatz for the radion potential:
\beq
	V(\varphi) = {\lambda\over4} \varphi^4 \left[ \left(
	\left({\varphi/f}\right)^{\epsilon} - \eta \right)^2
	+ (\alpha\eta\epsilon)^2 \right]
\eeq
Note that $\varphi = f e^{-r_1/L}$ is the radion field, not to be
confused with the bulk scalar $\phi$.   In this notation, the small
parameter
$\epsilon$ is related to our superpotential parameters by $\epsilon = 
bL$, $\eta = \phi_1/\phi_0$, the quartic coupling is 
$\lambda\cong 4\phi_0^2 / (9 L M_p^4)$, and 
$f \cong \sqrt{6}M_p$ \cite{CF1}.  
When $\alpha=0$, this agrees with the approximate form found by
ref.\ \cite{GW2} in the case where $\Lambda = 0$.  It has two degenerate
minima at $\varphi=0$ and $\varphi = f\eta^{1/\epsilon}$, corresponding
to the solutions with infinite and finite interbrane distance,
respectively, which we found above.  (This can be seen through the 
relation $\varphi/f = e^{-r_1/L}$ between the radion field and the 
brane separation.)  
By adding the term $(\alpha\eta\epsilon)^2$, we have lifted the minimum near 
$\varphi \cong f \eta^{1/\eps}$ to a nonzero value of the 4-D cosmological
constant:
\beq
	 \Lambda = {\lambda\over 32}\, \varphi_+^4 \eta^2 \epsilon^2
	\delta
\eeq
where in the limit that $\epsilon\ll 1$, the value of $\varphi$ at the
metastable minimum is given by
\beq
	\varphi_+ \cong f \eta^{1/\epsilon} e^{-\delta/4}; \qquad
	\delta \equiv 1 - \sqrt{1 - (4\alpha)^2}
\eeq

To estimate the bounce action, we use the thin wall approximation of
ref.\ \cite{Coleman}.  For a bubble of radius $R$ in 4-D Euclidean space,
with false vacuum energy density $\Lambda$, 
\beq
\label{Sb}
	S_b = -{\pi^2\over 2} \Lambda R^4 + 2\pi^2 R^3 S_1
\eeq 
where $S_1$ is the action for the 1-D instanton corresponding to the
$\alpha=\Lambda=0$ limit of the radion potential,
\beq
	S_1 = \int_0^{\varphi_+} d\varphi \, \sqrt{2 V}
	  =\frac{1}{9} \sqrt{\frac{\lambda}{2}} \,\epsilon\eta 
          (f\eta^{1/\epsilon})^3,
\eeq
and the radius of the bounce solution which minimizes (\ref{Sb}) is
$R = 3 {S_1\over\Lambda}$. Substituting this value into the (4-D) Euclidean action, we obtain
\beq
	S_b \cong {17 \pi^2 e^{3\delta} \over \epsilon^2 \eta^2 \lambda
	\delta^3}
\eeq
which is so large for the parameters of interest for getting 
a large hierarchy of scales ($\epsilon\sim 0.01$) that the false vacuum
state is easily more long-lived than the present age of the universe.

The 1-D instanton obeys the equation of motion
$\ddot\varphi = {dV\over d\varphi}|_{\alpha=0}$,
which can be solved in terms of the Lerch transcendental function $\Phi$,
\beq
	\epsilon \eta \sqrt{\lambda/2}\, t = {1\over\varphi}
	\Phi\left( \eta^{-1}(\varphi/f)^\epsilon,\, 1,\, -1/\epsilon\right),
\eeq
as illustrated in figure 2.  (We take $\epsilon=0.011$ since
$\Phi(x,1,-1/\epsilon)$ is singular when $1/\epsilon$ is a positive
 integer.)
This has the desired behavior that
$\varphi\to
0$ as $t\to-\infty$ and $\varphi\to\varphi_+$ as $t\to\infty$.

\centerline{\epsfxsize=3.0in\epsfbox{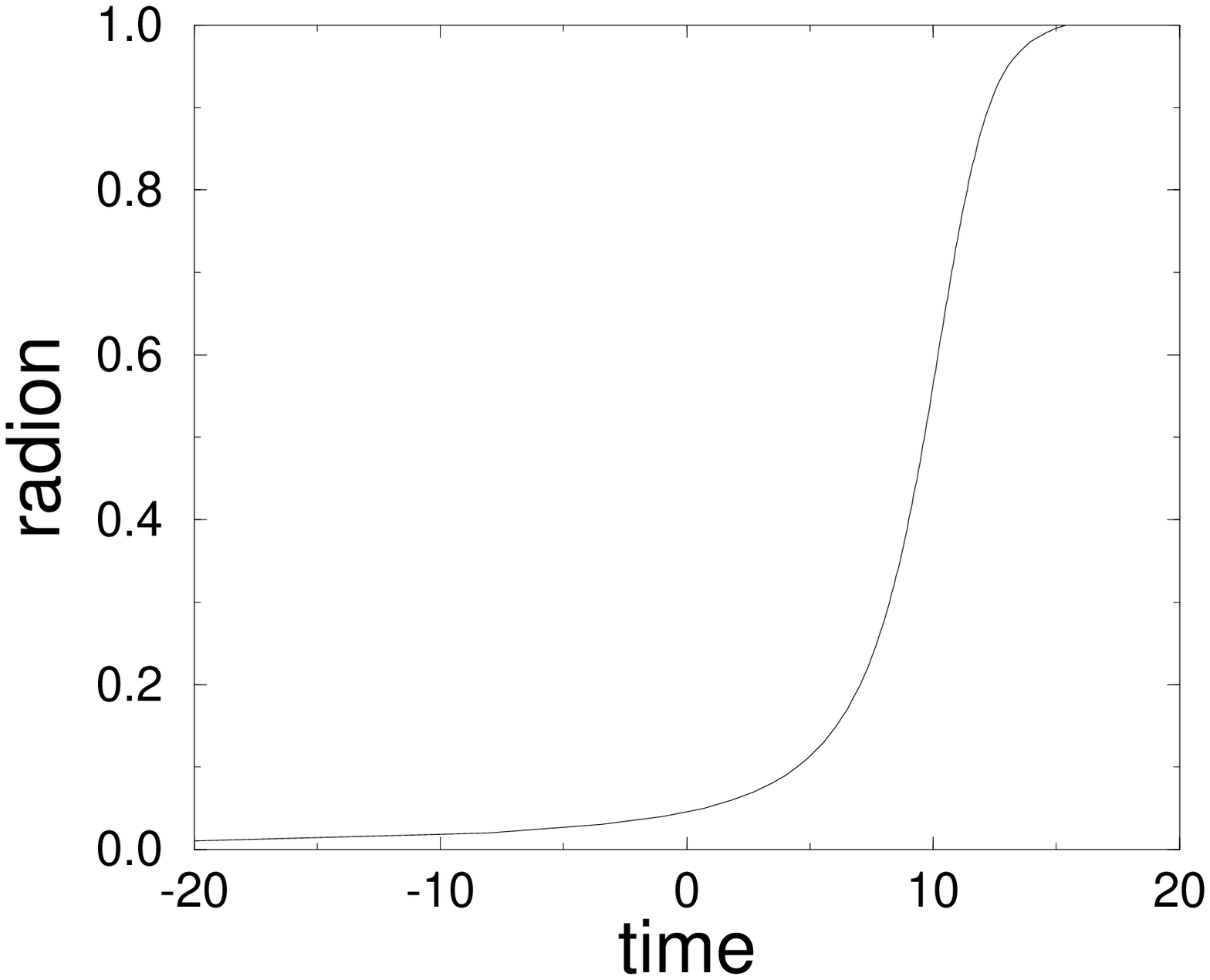}}
\vspace{0.in}
\noindent {\small
Figure 2: 1-D instanton solution for $\epsilon=0.011$ and $\delta=0$.  The axes are
scaled such that ``time'' $= \epsilon f \eta^{1+1/\epsilon} \sqrt{\lambda/2}
\,t$ and ``radion'' $= (\varphi/f)\eta^{-1/\epsilon}$.}
\bigskip

We can verify that the thin-wall approximation is valid when $\delta\ll
1$. From figure 2, the width is $\Delta R = \Delta t \cong 10 
(\epsilon f \eta^{1+1/\epsilon} \sqrt{\lambda/2})^{-1}$.
Comparing to the bounce radius
$R = 3S_1/\Lambda$, we find that $\Delta R/R \cong \delta$.

In ref.\ \cite{CF1} we studied a similar problem, that of transitions from a
false to a true minimum of the radion potential in the original
Goldberger-Wise mechanism.  There it was necessary to consider
thermally-induced transitions because the radion would have been in thermal
equilibrium  in the early universe.  In the present case, assuming that
inflation is driven by an inflaton on the observable brane, the
radion is so weakly coupled that it does not come into thermal
equilibrium.  This is also true of the variant model we propose below in
which the observable brane is at the TeV rather than the Planck scale.
\bigskip

{\bf 6.\ Physical consequences.}  In the false vacuum state, the radion
will
have a very small mass, somewhat below the milli-eV range, since it has
been shown that \cite{CGK}
\beq
\label{radion_mass}
	m_r \cong \sfrac83 b e^{-r_\sL/L}.
\eeq
Since $M_p e^{-r_\sL/L}$ is supposed to be the meV scale and $b\sim 0.01/L
\lsim 0.01 M_p$, this is a dangerously small mass if the radion were to 
couple to matter as strongly as does gravity.
We have investigated the corrections to this formula due to the
$O(\Lambda)$ perturbations and found that they are of the same
order as the unperturbed value if $\Lambda\sim b^2$.  Hence there is the
possibility that corrections to the radion mass which we cannot compute
by perturbing in $b$ make it somewhat larger than the value
(\ref{radion_mass}).

However, if we live on the positive tension (Planck) brane, the radion's
exponentially small couplings to matter make it impossible to observe
directly.  Let us denote the renormalized trace of the stress energy
tensor by $(\tilde T^\mu_\mu)_i = e^{-4r_i/L} (T^\mu_\mu)_i$ at the two branes
located respectively at $r=r_0=0$ and $r=r_1=r_\sL$.  Because the
radion wave function grows like $e^{2r/L}$ away from the Planck brane
\cite{CGR}, its coupling to matter on the respective branes goes like 
\cite{CGRT, GW3}
\beq
 	{\cal L}_{\rm radion} = {\varphi\over \Lambda^{1/4}}\left( (\tilde
T^\mu_\mu)_{_1} + 
	e^{-2r_1/L} (\tilde T^\mu_\mu)_{_0} \right).
\eeq
(In the original RS model, the TeV scale appeared in place of the
milli-eV scale $\Lambda^{1/4}$.)
If we are living on the Planck brane and $(\tilde T^\mu_\mu)_{_0}$
represents standard model physics, then the radion coupling to
the standard model is negligible since $e^{-2r_1/L} \Lambda^{-1/4}\sim
\Lambda^{-1/4}/M_p^2$.

There is a slightly better chance of detecting the effects of such a
radion through cosmology.  The shift in size of the extra dimension due to
physical energy density $\rho$ on the Planck brane is \cite{CGK,CGRT,CF2}
\beq
	{\Delta r_1\over r_1}
	 \cong {L \rho \over 6 r_1 m_r^2 M_p^2}
\eeq
which becomes of order unity at temperatures $T\sim \sqrt{m_r M_p}\sim
1$ TeV.  However this is still such a high temperature that it is
difficult to imagine any surviving remnant of the changes to physical
scales and the Hubble expansion rate which would arise from a variation
in $r_1$ during this era.

A more testable and interesting situation is to imagine that we are living
on a third brane located approximately halfway between the original two,
instead of on the positive tension brane.  This is desirable apart from
considerations of the cosmological constant problem, in that it preserves
the natural resolution of the weak scale hierarchy which was the original
motivation of Randall and Sundrum.  An intermediate brane can be inserted
into our solution if it has zero tension and a potential of the form
$\gamma_{1\over 2}(\phi-\phi_{1\over 2})^2$, with $\gamma_{1\over
2}\to\infty$ for ease of analysis.  The position of this new TeV brane can
be adjusted to the desired value by satisfying the approximate boundary
condition $e^{-br_{1/2}}\cong \phi_{1\over 2}/\phi_0$.  There are now two
radions, one at the TeV scale associated with fluctuations in the distance
between the Planck and TeV branes, and the original milli-eV radion.

If there does exist a TeV brane, the coupling of the light radion to
particles there will be significantly larger than on the Planck brane:
\beq
	{\cal L}_{\rm radion-\atop TeV\ brane} \cong {\varphi\over
 	\varphi_+}
%	\Lambda^{1/4}}
	\, e^{2(r_{1/2}-r_1)/L} \, (\tilde T^\mu_\mu)_{1\over 2} \sim
 	{\varphi\over M_p} (\tilde T^\mu_\mu)_{1\over 2},
\eeq
which means the radion is coupled with approximately the same strength as
ordinary gravity.  This is precisely the range of scalar masses and
couplings which is presently being probed by measurements of the
gravitational force at submillimeter range \cite{Eotwash}.  The new
contribution to the potential energy for two masses $m_1$ and $m_2$
separated by a distance $r$ is
\beq
	\Delta V = -{ 4\pi G_N m_1\, m_2\,
	e^{-m_r r}\over 3 r}\, e^{(4 r_{1/2}- 2 r_1)/L}
\eeq
The exponential factor should be of order (meV)$^2 M_p^2 /$(TeV)$^4 \sim
6$ for the warp factors to naturally explain the TeV and meV scales of the
standard model and cosmological constant, respectively.  
To evade the constraints of submillimeter gravity tests, this number must
be made somewhat smaller, or the radion mass must be made somewhat larger
than 1 meV.  The model is therefore tightly constrained by present tests
of the gravitational force.

In addition, the Kaluza-Klein gravitons have a mass gap similar in size to
the meV radion, and to the extent that they are nearly massless, their
effects will be like those of the KK gravitons in the noncompact
Randall-Sundrum scenario \cite{RS2}. We find that the correction to the
Newtonian gravitational potential from these modes is given approximately 
by
\beq
\Delta V \cong -{ G_N L^2 m^2} { m_1\, m_2\over r}\, {e^{-5mr/4}\over 1 - 
e^{-mr}}\, \left(\frac14 + {1\over 1-e^{-mr}}\right)
\eeq
where $m\equiv \pi e^{-r_1/L}/L$, and we have approximated the KK masses by
$m_n \cong (n+1/4) m$ using the large-$n$ behavior of the exact
eigenvalues \cite{DHR}.
Moreover, the existence of the TeV brane would imply the usual TeV radion,
whose phenomenology has been widely studied in connection with the compact
Randall-Sundrum scenario \cite{CGRT, GW3, radion}.
\bigskip

{\bf 7.\ Conclusions.} We have presented a warped compactification model
which, at the expense of assuming there is some unknown solution to the
first cosmological constant problem---the question of why the ultimate
vacuum energy of the universe is zero---naturally resolves the second one:
it explains how the observed value can be 120 orders of magnitude below
the Planck energy density without requiring additional fine tuning.  The
hypothesis is that there exists a brane at such a distance ($\sim
70/M_p$) from the Planck brane that the mistuning of its tension from the
flat-brane value contributes to $\Lambda$ at the $10^{-120} M_p^4$ level,
due to the smallness of the warp factor.

Our idea has several shortcomings.  Unlike quintessence models, it does
not try to solve the coincidence problem of why $\Lambda$ happens be a
significant fraction of the critical density {\it now}.  In the version
where we are assumed to live on the Planck brane, we sacrifice any new
understanding of the weak scale hierarchy problem, and furthermore there
seems to be no experimental signatures that would test the idea.  These
difficulties are overcome if we live instead on a zero-tension TeV brane
between the two original branes, for then the light radion is in the right
range for current tests of gravity at submillimeter distances, but then a
new fine-tuning problem is introduced: why is the tension exactly (or so
nearly) zero?  We nevertheless feel that the idea is worth pointing out,
in the hope that the problems might be surmounted.  For example, ref.\
\cite{KT} has advocated a self-tuning solution \cite{self-tuning} to the
first cosmological constant problem which avoids the problems of
singularities by letting a kink in the scalar field play the role of the
Planck brane instead of inserting it by hand.  Perhaps such an approach
can be used to eliminate some of the tunings that are still present in
ours.  We find it intriguing that this explanation of the cosmological
constant, whose presence is revealed by gravity over cosmological distance
scales, might be corroborated by table-top gravity experiments, as well as
collider searches.
\bigskip

We thank Richard MacKenzie and Michael Turner for discussions during the
early stages of this work, and Stephan Huber for useful comments.

\newpage
%%%%%%%%%%%%%%%%%%%%%%%%%%%%%%%%%%%%%%%%%%%%%%%%%%%%%%

\end{document}